\documentclass[twocolumn,showkeys,showpacs,preprintnumbers,prd,superscriptaddress,nofootinbib]{revtex4-1}
\usepackage{graphicx}	
\usepackage{amssymb}
\usepackage{dcolumn}
\usepackage{mathtools}
\usepackage{amsmath}
\usepackage{xcolor}
\usepackage{color}
\usepackage{theorem}
\usepackage{subfigure, rotating, bm, array}
\usepackage[pagebackref=false, colorlinks=true]{hyperref}
\hypersetup{linkcolor=blue, citecolor=blue,urlcolor=blue}

\begin{document}

\title{Tidal forces in the Simpson-Visser black-bounce and wormhole spacetimes}

\author{Dhruv Arora}
\email{arora09dhruv@gmail.com}
\affiliation{PDPIAS, Charusat University, Anand-388421 (Gujarat), India}

\author{Parth Bambhaniya}
\email{grcollapse@gmail.com}
\affiliation{PDPIAS, Charusat University, Anand-388421 (Gujarat), India}
\affiliation{International Center for Cosmology, Charusat University, Anand-388421 (Gujarat), India}

\author{Dipanjan Dey}
\email{deydipanjan7@gmail.com}
\affiliation{Department of Mathematics and Statistics,
Dalhousie University,
Halifax, Nova Scotia,
Canada B3H 3J5}

\author{Pankaj S. Joshi}
\email{psjcosmos@gmail.com}
\affiliation{International Centre for Space and Cosmology, School of Arts and Sciences, Ahmedabad University, Ahmedabad-380009 (Gujarat), India}
\affiliation{International Center for Cosmology, Charusat University, Anand-388421 (Gujarat), India}

\date{\today}

\begin{abstract}
The concept of regular black holes has gained attention in recent years, especially in the context of quantum gravity theories. In these theories, the existence of singularities is paradoxical as they represent a breakdown of the laws of physics. Motivated by the recent developments in this area, we study the tidal force effects in one such family of regular geometries described by the Simpson-Visser metric. We find the radial and angular force profiles for a radially in-falling particle in this spacetime and calculate the variation of the geodesic separation vector with the radial coordinate using two different initial conditions. These results are then compared with that of Schwarzschild black hole spacetime. We show that for a regular black hole, both radial and angular tidal forces show a peak outside the horizon and then fall to ultimately switch their behavior from stretching to compression and vice-versa. Also, they are finite at $r=0$ unlike the Schwarzschild spacetime. It is also seen that the angular deviation profile shows an oscillating behavior for a particular initial condition.  Our analysis can be used to distinguish between regular black hole, one-way and two-way wormholes and a singular black hole spacetimes.

\begin{description}
\item[Keywords]
Regular black holes, Tidal force, Wormholes
\end{description}
\end{abstract}

\maketitle


\section{\label{sec:level1}Introduction}

Various observations have revealed the general relativistic effects of Einstein's gravity, including the great discoveries of gravitational waves \cite{LIGOScientific:2016aoc}, the first ever black hole shadow images of the M87 and Milky Way's galactic centre \cite{EventHorizonTelescope:2022xnr,EventHorizonTelescope:2022tzy,EventHorizonTelescope:2022ago,EventHorizonTelescope:2022vjs,EventHorizonTelescope:2022wok,EventHorizonTelescope:2022urf,EventHorizonTelescope:2022xqj,EventHorizonTelescope:2022gsd}, stellar motions of S-stars around the Milky Way galactic centre \cite{Do:2019txf,GRAVITY:2018ofz,GRAVITY:2020gka}, and so on. These discoveries have sparked a lot of interest in researching the
the causal structure of spacetime and its dynamics around the galactic centre.

Despite the fact that black holes were predicted by GR more than a century ago as a solution to Einstein's equations, the physical properties of black holes remain very interesting to the scientific community. Many solutions to Einstein's equations followed, generalising Schwarzschild solutions: Kerr, Reissner-Nordstrom and Kerr-Newman black holes.

However having these achievements, there are still some conceptual issues with black holes in general relativity. The existence of physical singularity is the main problem associated with black holes. These issues a still unresolved as of today. Many physicists have speculated that a quantum gravity theory could solve this problem. However, even after the EHT findings, the phenomenology of black holes remains a hot topic. Regular black holes, wormholes, naked singularities, grava-stars, and other black hole mimickers are also becoming more common in the literature \cite{Janis:1968zz,Joshi:2011zm,Morris:1988tu,Hawking:1988ae,Schunck:2003kk,Simpson:2018tsi,Moti:2021vck}.

All of the previously discovered black hole solutions have a curvature singularity, where geometrical and physical quantities diverge. 
J. M. Bardeen proposed a regular black hole solution and noted the existence of solutions free of curvature singularities  \cite{Rodrigues:2022qdp}. This Bardeen black hole is an exact solution of Einstein's equations coupled to nonlinear electrodynamics. One reason for looking into such compact objects is that they could be black hole alternatives. Furthermore, as is well known, many of the possible observational signatures of black holes, such as precession of timelike bound orbits, gravitational lensing, shadow properties, and energy extractions can be mimicked by black hole mimickers \cite{Liu:2021yev,Joshi:2013dva,Bambhaniya:2021ugr,Tahelyani:2022uxw,Rahaman:2021kge,Harko:2008vy,Harko:2009xf,Kovacs:2010xm,Guo:2020tgv,Chowdhury:2011aa,Shaikh:2019jfr,Shaikh:2018oul,Virbhadra:2007kw,Gyulchev:2008ff,Sahu:2012er,Martinez,tsirulev,Joshi:2019rdo,Bambhaniya:2019pbr,Dey:2019fpv,Bam2020,Bambhaniya:2022xbz,Vagnozzi:2019apd,Chen:2022nbb,atamurotov_2015,abdujabbarov_2015b,Vagnozzi:2022moj,Saurabh:2022jjv,Patel:2022jbk,Patel:2023efv, Dey:2020bgo, Dey:2020haf}. 

From a historical standpoint, Einstein and Rosen proposed the existence of "bridges" through space-time using general relativity theory \cite{Einstein:1935tc}. These bridges connect two points in space-time, allowing Einstein-Rosen bridges, also known as wormholes, to be created. Such wormholes have been shown to be non traversable. The concept of traversable wormholes was later explored in the seminal work of Morris and Throne \cite{Morris:1988cz}. As a result, we can use astrophysical data to test black holes and wormholes. The authors of Ref. \cite{Jusufi:2021lei} investigated the possibility of testing wormhole geometries in our galaxy using an in-falling and radiating gas accretion model and the motion of the S2 star. Refs. \cite{Dai:2019mse,Bambi:2021qfo} investigated the possibility of testing wormholes by observing the galactic centre using the in-falling gas model. References \cite{Simonetti:2020ivl} contain additional interesting works on the formation and stability of wormholes.

Simpson and Visser proposed \cite{Simpson:2018tsi} a very simple theoretically appealing spherically symmetric and static spacetime family derived from Schwarzschild geometry that allows for a unique description of regular black holes and wormholes via smooth interpolation between these two possibilities via a length-scale parameter that drives the regularisation of the central singularity. Its rotational form was also recently demonstrated \cite{Mazza:2021rgq}. Ref. \cite{Bronnikov:2021uta} shows that this spacetime can be obtained as a solution to Einstein field equations sourced by a combination of a minimally coupled phantom scalar field with a non-zero potential and a magnetic field in the framework on non linear electrodynamics. It is now well understood that a test body in free fall towards the centre of another massive body is stretched in the radial direction while compressed in the angular direction. Gravity's tidal effect, which is caused by a variation in gravity's strength between two adjacent points, causes the stretching and compression. Tidal force phenomena are extremely common in our universe and have been the subject of popular scientific investigation for much of the twentieth century.

The authors of \cite{Crispino:2016pnv} investigated the effects of tidal force on a freely falling particle in a Reissner-Nordstrom (RN) black hole and discovered that the radial and angular components of the tidal force change signs at a boundary known as the event horizon (a null hypersurface), which is not the case in a Schwarzschild black hole.
The authors of \cite{Gad:2010ion} compared the tidal force effects in a stringy charged black hole to those in Schwarzschild and RN black holes. The radial and angular components of the tidal force in the Kiselev black hole were studied by Shahzad and Jawad \cite{Shahzad:2017vwi}. Motivated by the above developments, in this paper we study the tidal force effects and geodesic deviation in the Simpson-Visser spacetime. We then compare our results with the Schwarzschild case to look for and study any significant changes that can shed some light on the nature of black hole mimickers.

This paper's overview is as follows: In Section (\ref{sec:II}), we discussed the features of this spacetime, i.e. a synopsis of previous work on this metric and its properties. We present the calculations for tidal force effects for a radially in-falling particle in this spacetime in section (\ref{sec:III}) and obtain radial and angular force profiles using tetrad formalism. Radial and angular force profiles and their behavior is discussed in section (\ref{IV}). We compared the numerical solution of the geodesic deviation equation for the Simpson-Visser metric with the Schwarzschild black hole metric in section (\ref{V}). Conclusions are given in section (\ref{VI}). Throughout the paper, we have used gravitational constant ($G$) and velocity of light ($c$) as unity.

\section{\label{sec:II}Simpson-Visser SPACETIME}
The Simpson-Visser metric represents a spherically symmetric and static class of black hole mimickers with a minimal surface in place of the central singularity. It is defined by the metric \cite{Simpson:2018tsi}:
\begin{equation}
   ds^{2} = -f(r)dt^{2}+f(r)^{-1}dr^{2}+(r^{2}+l^{2})(d\theta^{2}+\sin^{2}\theta d\phi^{2})
   \label{1}
\end{equation}

where \begin{eqnarray}
    f(r) = 1-\frac{2M}{\sqrt{r^{2}+l^{2}}}
    \label{2}
\end{eqnarray}
In the above equation, $M \ge 0$ is the ADM mass and $l > 0$ is the regularization parameter of the central singularity, and possibly reflects quantum gravity effects. This spacetime is carefully designed to be the minimalist modification of the Schwarzschild spacetime. By varying the parameter $l$, the metric can represent the following cases:

\begin{enumerate}
    \item The ordinary Schwarzschild spacetime ($l=0$).
    \item A regular black hole geometry with one-way spacelike throat ($l <2M$).
    \item A one-way wormhole geometry with external null throat ($l = 2M$).
    \item A traversable wormhole geometry with two-way timelike throat ($l> 2M$).
\end{enumerate}
The geometry in the region where it represents a regular black hole is unusual in that it describes a bounce into a future incarnation of the universe rather than a bounce back into our own \cite{Barcelo:2014cla,BenAchour:2020gon}. This metric does not correspond to a traditional regular black hole geometries such as the Bardeen, Bergmann-Roman, Frolov, or Hayward \cite{Hayward:2005gi,Frolov:2017rjz,Cano:2018aod,Bardeen:2018frm}. Instead, it is either a regular black hole (bouncing into a future incarnation of the universe) or a traversable wormhole, depending on the value of the parameter $l$. Latest studies done by authors in \cite{Zhou:2022yio}  argue on the geodesic incompleteness of some popular regular black holes such as analytically extended Hayward black hole and the extension of the Culetu-Simpson-Visser’s non-analytic smooth black hole. Similar analysis can also be extended to the metric given in Eq. (\ref{1}) to test its geodesic completeness.
The coordinate location of the horizons in this spacetime when using Schwarzschild coordinates ($t$,$r$,$\theta$,$\phi$) is given by imposing $g^{rr} = 0$ which gives us:
\begin{eqnarray}
 r_\pm = \sqrt{(2M)^{2} - l^{2}}.   
 \label{3}
\end{eqnarray}
Also, by studying the radial null curves in this geometry i.e. setting $ds^{2} = 0$, $d\theta = d\phi =0$ we get:
\begin{eqnarray}
    \frac{dr}{dt} = \pm\bigg(1-\frac{2M}{\sqrt{r^{2}+l^{2}}}\bigg).
    \label{4}
\end{eqnarray}
This equation also defines the coordinate speed of light in this metric. For all values of $l < 2M$, we get a pair of horizons at symmetrically placed $r$ coordinate given by Eq. (\ref{3}). If $l = 2M$, then as $r \rightarrow 0$, we have $\frac{dr}{dt} \rightarrow 0$. This means there is a horizon at $r=0$ in the case of the one-way wormhole with an external null throat. If $l > 2M$, then  $\frac{dr}{dt} \ne 0$ for any value of $r$. This suggests that the geometry is a two-way traversable wormhole. 

We can use scalar curvature invariants to detect the event horizon for the above-mentioned Simpson-Visser spacetime. Describing the presence of a horizon using scalar curvature invariants for a given spacetime is very useful since this description is coordinate-independent \cite{Abdel,Coley:2017woz, mcnut}. We can write down the following dimensionless scalar invariant \cite{Abdel}:
\begin{eqnarray}
    \mathcal{Q}_2=\frac{1}{27}\frac{I_5~ I_6 - I_7^2}{\left(I_1^2 + I_2^2\right)^{\frac{5}{2}}}\,\, ,
\end{eqnarray}
where $I_5 = k_\mu k^\mu$ where $k_\mu = -\nabla_\mu I_1$, $I_6 = n_\mu n^\mu$ where $n_\mu = -\nabla_\mu I_2$, and $I_7 = k_\mu n^\mu$. Here, $I_1 = C_{abcd}C^{abcd}$ and $I_2 = C_{abcd}^*C^{abcd}$ where $C_{abcd},~ C_{abcd}^*$ are the Weyl curvature tensor and its dual tensor, respectively. The dimensionless scalar invariant $\mathcal{Q}_2$ becomes zero at $r=0$ and at $r=\sqrt{(2M)^{2} - l^{2}}~$ i.e., at the event horizon when $l<2M$. This is true for any values of angular co-ordinates, however, at $\theta = \frac{\pi}{2}$ plane, $\mathcal{Q}_2$ is identically zero $\forall r$. When $l>2M$, $\mathcal{Q}_2$ becomes zero at $r=0$ only which suggests the absence of the event horizon. Therefore, in Simpson-Visser spacetime, $\mathcal{Q}_2 = 0, \forall r>0, \forall \theta \neq \frac{\pi}{2}$ implies the presence of event horizon (i.e., a null hypersurface $\Sigma = \partial J^- (\mathcal{I}^+)\cap \mathcal{M}$, where $J^- (\mathcal{I}^+)$ is the causal past of the future null infinity $\mathcal{I}^+$) which is true in any coordinate system.

Another very crucial property of any spacetime that needs to be verified is whether its ADM mass is finite since it is related to the asymptotic flatness of that spacetime. An asymptotically Minkowskian spacetime would also be asymptotically flat if the ADM mass of that spacetime is finite. Spacetimes which are asymptotically Minkowskian but don't have finite ADM mass are known as quasi-flat spacetime \cite{ Sud}.  The ADM mass for a spacetime manifold ($\mathcal{M}, g_{\alpha\beta}$) can be written as:
\begin{equation}
M_{ADM} =   -\frac{1}{8 \pi} \lim_{S_{t} \to \infty} \oint_{S_{t}} (K-K_{0})\sqrt{q} \quad d^{2}y
\end{equation}
where $S_{t}$ is the bounded spacelike two-surface, $K$ is the extrinsic curvature of the two-surface $S_{t}$ embedded in the three-dimensional spacelike hypersurface $\Sigma_{t}$, $K_{o}$ is the extrinsic curvature of the two-surface $S_{t}$ embedded in the flat space, and the infinitesimal area of the two-surface is given by $\sqrt{q} ~d^{2}y = \left(r^{2} + l^2\right) sin \theta ~d\theta~ d\phi$, where $q$ is the determinant of the induced metric on the two-surface. 
The extrinsic curvature of $S_{t}$ for the Simpson-Visser space-time is:
\begin{eqnarray}
    K = \frac{2 r}{l^2 + r^2}\bigg(1-\frac{2M}{\sqrt{r^{2}+l^{2}}}\bigg)^{\frac12}\,\, ,
\end{eqnarray}
and $K_0 = \frac{2}{r}$. Therefore, the ADM mass $$M_{ADM}= \lim_{r \to \infty}\left[r\left(1+\frac{l^2}{r^2}\right)-r\sqrt{1-\frac{2M}{\sqrt{r^{2}+l^{2}}}}\right] = M,$$
which is finite. One can also verify that the Simpson-Visser spacetime is asymptotically Minkowskian, since $\lim_{r \to \infty} g_{tt} = -1,~ \lim_{r \to \infty} g_{rr} = 1$, and $\lim_{r \to \infty} g_{\theta\theta} = r^2, ~\lim_{r \to \infty} g_{\phi\phi} = r^2 \sin^2\theta$. Therefore, this space-time is asymptotically flat.

\begin{figure*}[htbp]
  \centering
  \begin{minipage}[b]{0.47\linewidth}
    \centering
    \includegraphics[width=\linewidth]{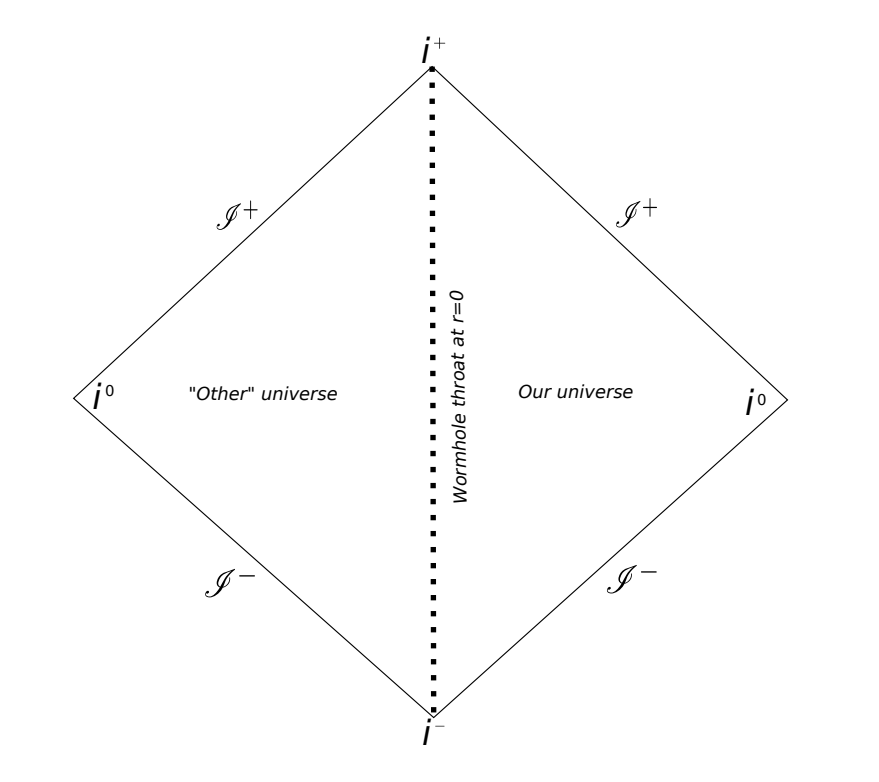}
    \caption{Penrose diagram for $l > 2M$ corresponding to a two-way wormhole \cite{Simpson:2018tsi}.}
    \label{fig:image1}
  \end{minipage}
  \hspace{0.5cm}
  \begin{minipage}[b]{0.48\linewidth}
    \centering
    \includegraphics[width=\linewidth]{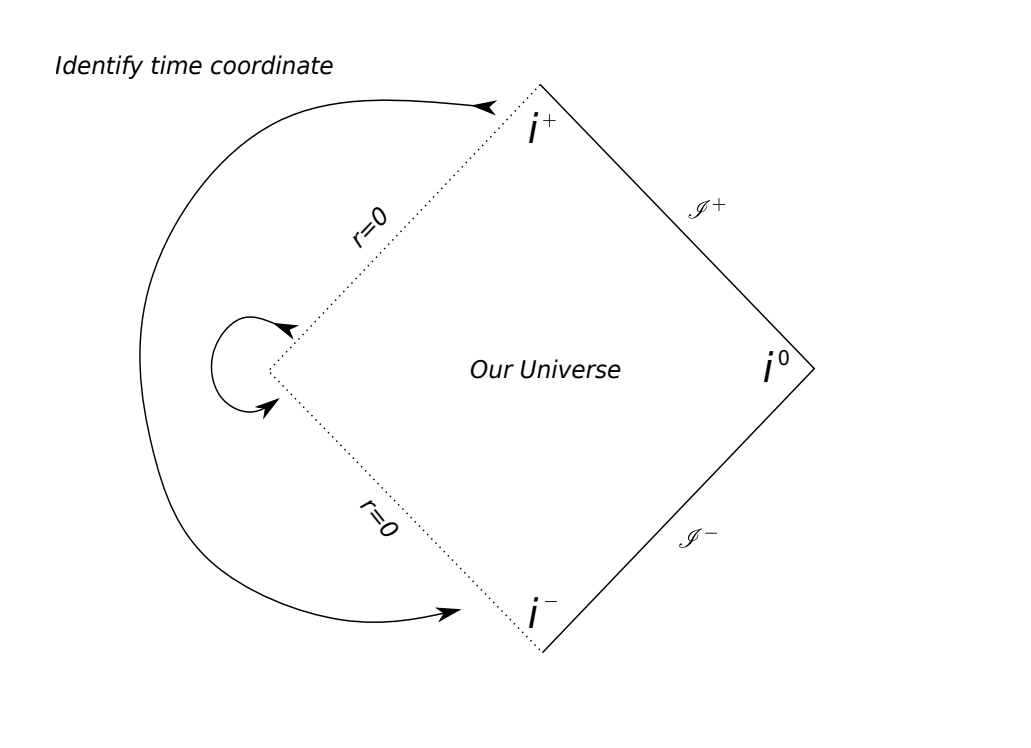}
    \caption{Penrose diagram for $l = 2M$ corresponding to a one-way wormhole \cite{Simpson:2018tsi}.}
    \label{fig:image2}
  \end{minipage}
\end{figure*}

\section{\label{sec:III}TIDAL FORCE EFFECTS}
In this section, we will delve into the complexities of the equations involving tidal forces in our spacetime. To investigate the equation for the distance between two infinitesimally close and free falling particles, we use the following equation for the spacelike components of the geodesic deviation vector $\eta^{\mu}$
\begin{eqnarray}
    \frac{D^{2} \eta^{\mu}}{D\tau^{2}} = R^{\mu}_{\alpha \beta \gamma} v^{\alpha}  v^{\beta} \eta^{\gamma},
    \label{5}
\end{eqnarray}
where $R^{\mu}_{\alpha \beta \gamma}$ is the Riemann curvature tensor and $v^{\mu}$ is the unit tangent vector to the geodesic. It is known that a test body moving in space solely under the influence of gravity follows a geodesic \cite{Falcon-Gomez:2022wze}. Now, due to the difference in curvature at every point on the geodesic and the fact that each point on the test body tends to follows a unique geodesic, the body experiences a difference in acceleration, leading to a stretching and squeezing effect known as tidal forces. This is the reason that the Riemann curvature tensor is used to study the tidal interactions due to gravity.

To calculate the tidal forces in the falling body's frame, we make use of the tetrad formalism \cite{Mitsou:2019nhj}. Tetrads are geometric objects that form a set of local coordinate bases, i.e. a locally defined set of four linearly independent vector fields known as tetrads or vierbien. At each point on a geodesic, there is a tetrad frame that forms a local inertial reference frame, where the laws of special relativity apply. Since, Lorentz transformations can connect an infinite number of orthonormal bases at a specific point, they cannot give enough information about the connection on the metric. To resolve this, we use the transformation from the orthonormal basis to the coordinate basis:
\begin{eqnarray}
    \overrightarrow{e}_{\mu} = \hat{e}^{\hat{\mu}}_\mu  \overrightarrow{e}_{\hat{\mu}},
    \label{6}
\end{eqnarray}
where $\overrightarrow{e}_{\mu}$ represents the coordinate basis, $\overrightarrow{e}_{\hat{\mu}}$ represents the orthonormal basis and $\hat{e}^{\hat{\mu}}_\mu $ are the tetrad components. The metric tensor components in the tetrad basis are given as \cite{Mitsou:2019nhj}:
\begin{eqnarray}
    g_{\mu\nu} = \eta_{\hat{\mu}\hat{\nu}} \hat{e}^{\mu}_{\hat{\mu} } \hat{e}^{\hat{\nu}}_\nu. 
    \label{7}
\end{eqnarray}
This is the key equation that enables us to use orthonormal basis in curved spacetime. The tetrad components for freely falling frames can be obtained from Eq. (\ref{4}) and Eq. (\ref{5}).
\begin{equation}
    \hat{e}^{\mu}_{\hat{0} } = \bigg\{\frac{E}{f(r)} , -\sqrt{E^{2}-f(r)}, 0, 0\bigg\} ,
    \label{8}
\end{equation}
\begin{equation}
    \hat{e}^{\mu}_{\hat{1} } = \bigg\{\frac{-\sqrt{E^{2}-f(r)}}{f(r)}, E, 0, 0\bigg\},
    \label{9}
\end{equation}
\begin{equation}
    \hat{e}^{\mu}_{\hat{2} } = \bigg\{0, 0, \frac{1}{r}, 0\bigg\}, 
    \label{10}
\end{equation}
\begin{equation}
    \hat{e}^{\mu}_{\hat{3} } = \bigg\{0 ,0, 0, \frac{1}{r\sin\theta}\bigg\}, 
    \label{11}
\end{equation}
where the tetrad components follow the following rule:
\begin{equation}
     \hat{e}^{\mu}_{\hat{\nu} }  \hat{e}^{\hat{\rho}}_{\mu}  = \delta^{\hat{\rho}}_{\hat{\nu}}.
     \label{12}
\end{equation}
We notice that the component $\hat{e}^{\mu}_{\hat{0} } = v^{\mu}$ is the tangent vector to the geodesic i.e. the four velocity of the particle. Also, the geodesic deviation vector follows the transformation rule from global coordinates to local orthonormal coordinates:
\begin{equation}
    \eta^{\mu} = \hat{e}^{\mu}_{\hat{\nu} } \eta^{\hat{\nu}}.
    \label{13}
\end{equation}

\begin{figure}[h]
  \centering
  \begin{subfigure}{}
    \centering
    \includegraphics[width=\linewidth]{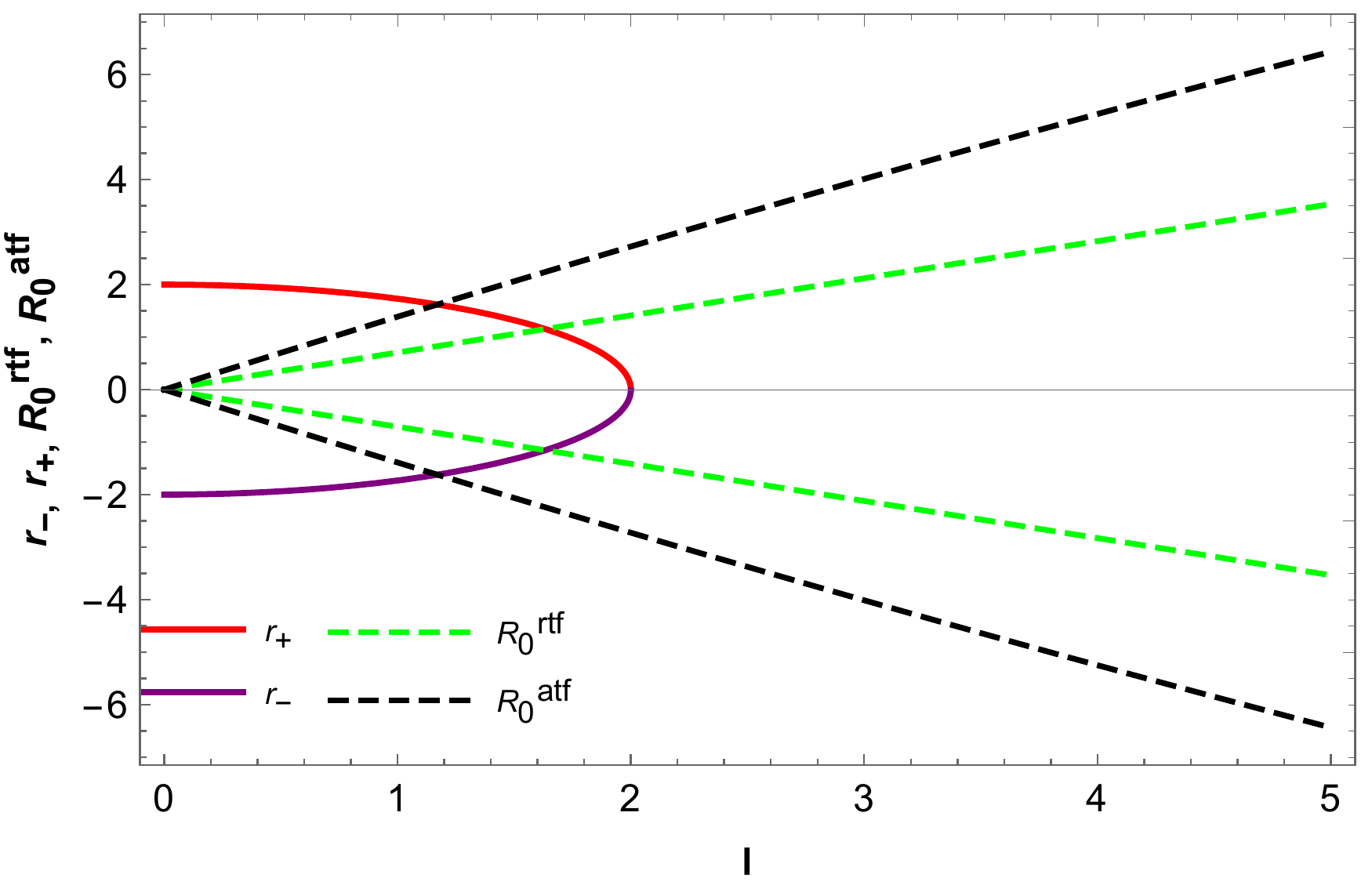}
    \caption{In this Figure, $r_{-}, r_{+}, R^{rtf}_0, R^{atf}_0$ plotted as a function of $l$. The intersection point of $r_{+}$ and $R^{rtf}_0$ happens at $\pm \frac{l}{\sqrt{2}}$, where the radial tidal force inverts its direction. Plots are made at a value of M = 1.}
  \end{subfigure}
  \label{fig:3}
  \end{figure}
The non vanishing independent components of Riemann tensor for spherically symmetric spacetimes, including Simpson-Visser are given by:
\begin{eqnarray}
   R^{1}_{2 1 2} = \frac{-r f(r)}{2} - f(r) +\frac{r^{2} f(r)}{r^{2}+l^{2}}, 
   \label{14}
\end{eqnarray}
\begin{eqnarray}
    R^{1}_{0 1 0} = \frac{f(r) f''(r)}{2},
    \label{15}
\end{eqnarray}
\begin{equation}
    R^{1}_{3 1 3} = \frac{-r f'(r)\sin^{2}\theta}{2} - \frac{l^{2} f(r)\sin^{2}\theta}{r^{2}+l^{2}},
    \label{16}
\end{equation}
\begin{eqnarray}
    R^{2}_{0 2 0} = \frac{r f(r) f'(r)}{2(r^{2}+l^{2})},
    \label{17}
\end{eqnarray}
\begin{eqnarray}
    R^{3}_{0 3 0} = \frac{r f(r) f'(r)}{2(r^{2}+l^{2})},
    \label{18}
\end{eqnarray}
\begin{eqnarray}
    R^{2}_{3 2 3} = \sin^{2}\theta - \frac{r^{2} f(r) \sin^{2}\theta}{r^{2}+l^{2}}.
    \label{19}
\end{eqnarray}
We see that substituting $l=0$ in the above equations yields the components for Schwarzschild spacetime. For computing the Riemann curvature tensor components in tetrad basis, we use the tetrad formalism as:
\begin{equation}
     R^{\hat{\mu}}_{\hat{\alpha} \hat{\beta} \hat{\gamma}} = R^{a}_{b c d} e^{\hat{\mu}}_a e^{b}_{\hat{\alpha}} e^{c}_{\hat{\beta}} e^{d}_{\hat{\gamma}}.
     \label{20}
\end{equation}
Following the Eq. (\ref{17}), the tidal tensor components are given by:  
\begin{eqnarray}
    R^{\hat{1}}_{\hat{0} \hat{1} \hat{0}} = \frac{f''(r)}{2},
    \label{21}
\end{eqnarray}
\begin{eqnarray}
   R^{\hat{2}}_{\hat{0} \hat{2} \hat{0}} =  \frac{rf'(r)}{2(r^{2}+l^{2})} + \frac{l^{2}{(f(r) - E^{2})}}{(r^{2}+l^{2})^{2}},
   \label{22}
\end{eqnarray}
\begin{eqnarray}
   R^{\hat{3}}_{\hat{0} \hat{3} \hat{0}} =  \frac{rf'(r)}{2(r^{2}+l^{2})} + \frac{l^{2}{(f(r) - E^{2})}}{(r^{2}+l^{2})^{2}}.
   \label{23}
\end{eqnarray}

\section{TIDAL FORCE EQUATIONS} \label{IV}
Upon obtaining the expressions from Eq. (\ref{18}-\ref{20}), we can obtain the relative acceleration between two nearby particles as:
\begin{eqnarray}
   \frac{D^{2}\eta^{\hat{r}}}{D\tau^{2}} =  \frac{-f''(r)}{2} \eta^{\hat{r}},
   \label{24}
\end{eqnarray}
\begin{eqnarray}
    \frac{D^{2}\eta^{\hat{\theta}}}{D\tau^{2}} = -\bigg[\frac{rf'(r)}{2(r^{2}+l^{2})}+l^{2}\bigg(\frac{f(r)-E^{2}}{(r^{2}+l^{2})^{2}}\bigg)\bigg] \eta^{\hat{\theta}},
    \label{25}
\end{eqnarray}
\begin{eqnarray}
    \frac{D^{2}\eta^{\hat{\phi}}}{D\tau^{2}} = -\bigg[\frac{rf'(r)}{2(r^{2}+l^{2})}+l^{2}\bigg(\frac{f(r)-E^{2}}{(r^{2}+l^{2})^{2}}\bigg)\bigg] \eta^{\hat{\phi}}.
    \label{26}
\end{eqnarray}
By substituting the values of $f(r)$ and its higher derivatives in Eq. (\ref{21}, \ref{22}, \ref{23}) we get:
\begin{eqnarray}
    \frac{D^{2}\eta^{\hat{r}}}{D\tau^{2}} = -\frac{M(l^{2}-2r^{2})}{(l^{2}+r^{2})^{5/2}} \eta^{\hat{r}},
    \label{27}
\end{eqnarray}
\begin{eqnarray}
    \frac{D^{2}\eta^{\hat{\theta}}}{D\tau^{2}} = \frac{-Mr^{2}+l^{2}(2M+(E^2-1)\sqrt{l^{2}+r^{2}})}{({l^{2}+r^{2})^{5/2}}} \eta^{\hat{\theta}},
    \label{28}
\end{eqnarray}
\begin{eqnarray}
   \frac{D^{2}\eta^{\hat{\phi}}}{D\tau^{2}} =  \frac{-Mr^{2}+l^{2}(2M+(E^2-1)\sqrt{l^{2}+r^{2}})}{({l^{2}+r^{2})^{5/2}}} \eta^{\hat{\phi}}.
   \label{29}
\end{eqnarray}

\subsection{RADIAL TIDAL FORCE}

\begin{figure}
    \centering
    \includegraphics[width = \linewidth]{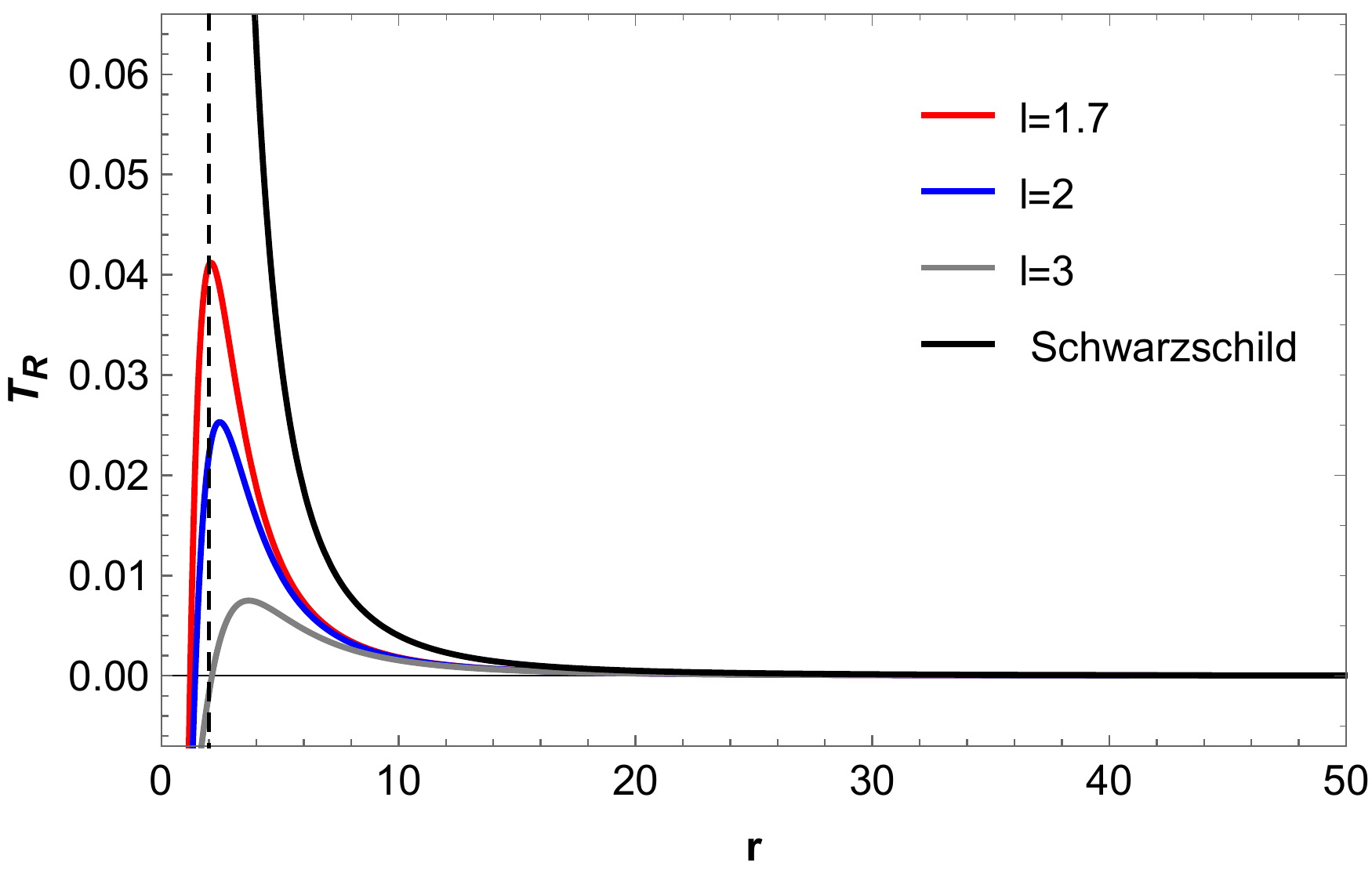}
    \caption{Radial tidal force as a function of the radial coordinate. We notice that the force vanishes at a single point in contrast to what happens with Schwarzschild spacetime. We take $M=1$ and $b=50 M$. Dotted line shows the location of horizon for Schwarzschild spacetime}
    \label{fig:4}
\end{figure}

We notice from Eq. (\ref{27}) that the radial tidal force reaches a finite value at $r = 0$, being equal to
\begin{eqnarray}
    \frac{D^{2}\eta^{\hat{r}}}{D\tau^{2}} \bigg |_{r=0} = \frac{-M}{l^{3}} \eta^{\hat{r}}.
    \label{30}
\end{eqnarray}
This is in contradiction to the Schwarzschild case where the radial tidal force tends to infinity (Fig 4). Also from Eq. (\ref{27}) it can be shown that the radial tidal force vanishes at a point $R = R_0^{rtf}$ given by:
\begin{eqnarray}
   R_0^{rtf} = \pm \frac{l}{\sqrt{2}}  .
   \label{31}
\end{eqnarray}
Radial tidal forces do not vanish in Schwarzschild spacetime, but are seen in Reissner-Nordstorm and other regular black hole spacetimes. By comparing this and Eq. (\ref{3}) for the location of event horizon we find that at $l = \pm 2\sqrt{6}$ the radial tidal force inverts its direction and becomes compressive at the event horizon. The radial tidal force takes a maximum value at $R^{rtf}_{max}$ given as:
\begin{eqnarray}
   R^{rtf}_{max} = \pm \sqrt{\frac{3}{2}}  l .
   \label{32}
\end{eqnarray}
It can be observed from the plot in fig (\ref{fig:4}) that the peak value of force shifts towards the left as the particle is made to radially fall from infinity. This suggests that in case of a regular black hole geometry, maximum radial stretching is achieved much closer to the horizon when compared with one-way and two-way wormhole geometries.

\subsection{ANGULAR TIDAL FORCE}
\begin{figure}[t]
    \centering
    \includegraphics[width = \linewidth]{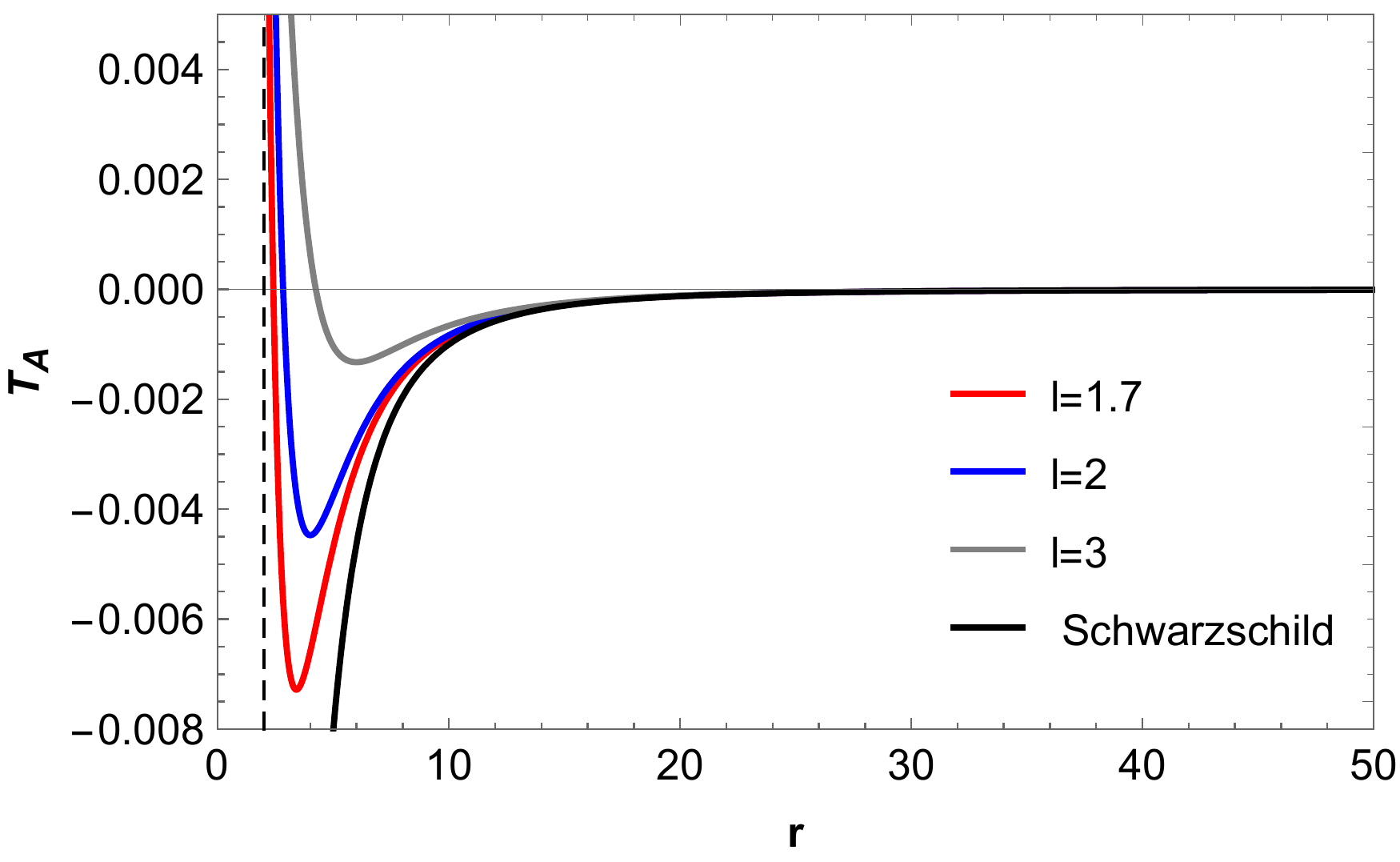}
    \caption{Angular tidal force as a function of the radial coordinate. We notice that the force vanishes at a single point in contrast to what happens with Schwarzschild spacetime. We take $M=1$, $E=1$ and $b=50 M$. Dotted line shows the location of horizon for Schwarzschild spacetime}
    \label{fig:5}
\end{figure}

Just like the radial tidal force , the angular tidal force also reaches a finite value at $r=0$ given by the solution of Eq.(\ref{28}) at as:
\begin{eqnarray}
   \frac{D^{2}\eta^{\hat{i}}}{D\tau^{2}} \bigg |_{r=0}  = \frac{2M+(E^{2}-1)l}{l^{3}} \eta^{\hat{i}}.
   \label{33}
\end{eqnarray}
where $i = \{\theta,\phi\}$. This result differs from the Schwarzschild case where the angular tidal forces diverge at $r=0$. Also, from Eq. (\ref{28}) the radial coordinate where the angular tidal force switches its sign can be found. The expression obtained is a sixth order polynomial, but can be simplified if we assume a numerical value of energy ($E$). In this study we use $E=1$ which results in the following value of $R_0^{atf}$:
\begin{eqnarray}
  R_0^{atf} = \pm\sqrt{2} l.
  \label{34}
\end{eqnarray}
It can be deduced from the graph that for a given value of energy, as the value of $l$ is increased, there comes a point at which the angular tidal force becomes entirely of stretching nature.Plot shown in fig(\ref{fig:5}) also shows peak value of force shifting to left for different geometries, suggesting the same conclusion as made in the case of radial tidal force.

\section{GEODESIC DEVIATION VECTOR}\label{V}
In this section we discuss the variation of geodesic deviation vector with the radial coordinate in Simpson-Visser spacetime. The deviation vector measures the deformation of a body falling radially in the Simpson-Visser spacetime. We can convert Eq. (\ref{27}) and (\ref{28}) in second derivatives w.r.t. $r$ by substituting $dr/d\tau$ = $-\sqrt{E^{2}-f(r)}$ which results from Eq. (\ref{8}). This gives us the following second order differential equations in $r$ as:
\begin{eqnarray}
  ( E^2-f(r)) \frac{D^{2}\eta^{\hat{r}}}{D r^{2}} - \frac{f'(r)}{2} \frac{D \eta^{\hat{r}}}{dr}+\frac{f''(r)}{2} \eta^{\hat{r}} = 0,
  \label{35}
\end{eqnarray}
\begin{eqnarray}
\begin{split}
   ( E^2-f(r)) \frac{D^{2}\eta^{\hat{i}}}{D r^{2}} - \frac{f'(r)}{2} \frac{D \eta^{\hat{i}}}{dr} +\bigg[\frac{r f'(r)}{2(r^{2}+l^{2})} + \\ l^{2} \bigg(\frac{f(r) - E^{2}}{(r^{2}+l^{2})^{2}}\bigg)\bigg] \eta^{\hat{i}} =0 .
\end{split}
\label{36}
  \end{eqnarray}
where $i = \{\theta,\phi\}$. The analytic solution for Eq. (\ref{34}) as pointed out in \cite{Lima:2020wcb} can be given as:
\begin{eqnarray}
  \eta^{\hat{r}} (r) = \sqrt{E^{2}-f(r)}\bigg[C_{1}+C_{2}\int \frac{dr}{(E^{2}-f(r))^{3/2}}\bigg] .
  \label{37}
\end{eqnarray}
where $C_{1}$ and $C_{2}$ are constants of integration. A similar solution can also be obtained for Eq. (\ref{35}). In order to find their value, we numerically solve the differential equations by imposing some initial conditions. For the purpose of this study we take the following initial conditions.
\begin{eqnarray}
    \eta^{\hat{\beta}} (b) >0 , \dot{\eta}^{\hat{\beta}} (b) = 0  ,\quad (ICI) 
    \label{38}
\end{eqnarray}
\begin{eqnarray}
   \eta^{\hat{\beta}} (b) =0 , \dot{\eta}^{\hat{\beta}} (b) > 0 .\quad (ICII)
   \label{39}
\end{eqnarray}
where $\beta = \{r,\theta,\phi\}$. $\eta^{\hat{\beta}} (b)$ represents the separation between two nearby geodesics at $r=b$ in the radial and angular directions. The initial condition in Eq. (\ref{37}) represents a mass released from rest at $r=b$ and the initial condition in Eq. (\ref{38})  a body constituted of dust to explode at $r=b$ . In the next sections we will discuss the components of radial and the angular deviation vectors in detail.

\subsection{RADIAL COMPONENT}
\begin{figure}[p]
  \centering
  \begin{subfigure}{}
    \centering
    \includegraphics[width=\linewidth]{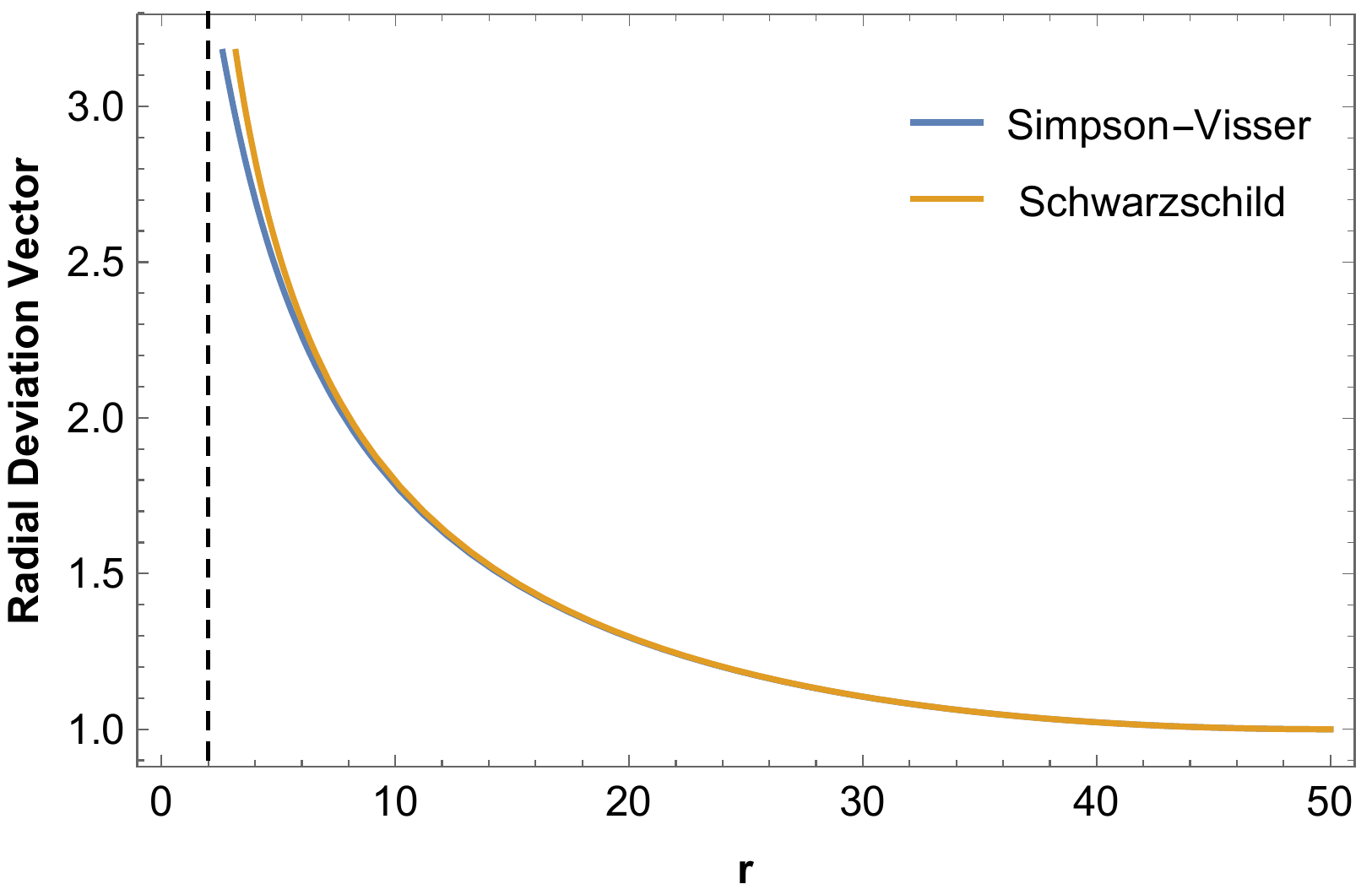}
    
  \end{subfigure}
 
 \begin{subfigure}{}
    \centering
    \includegraphics[width=\linewidth]{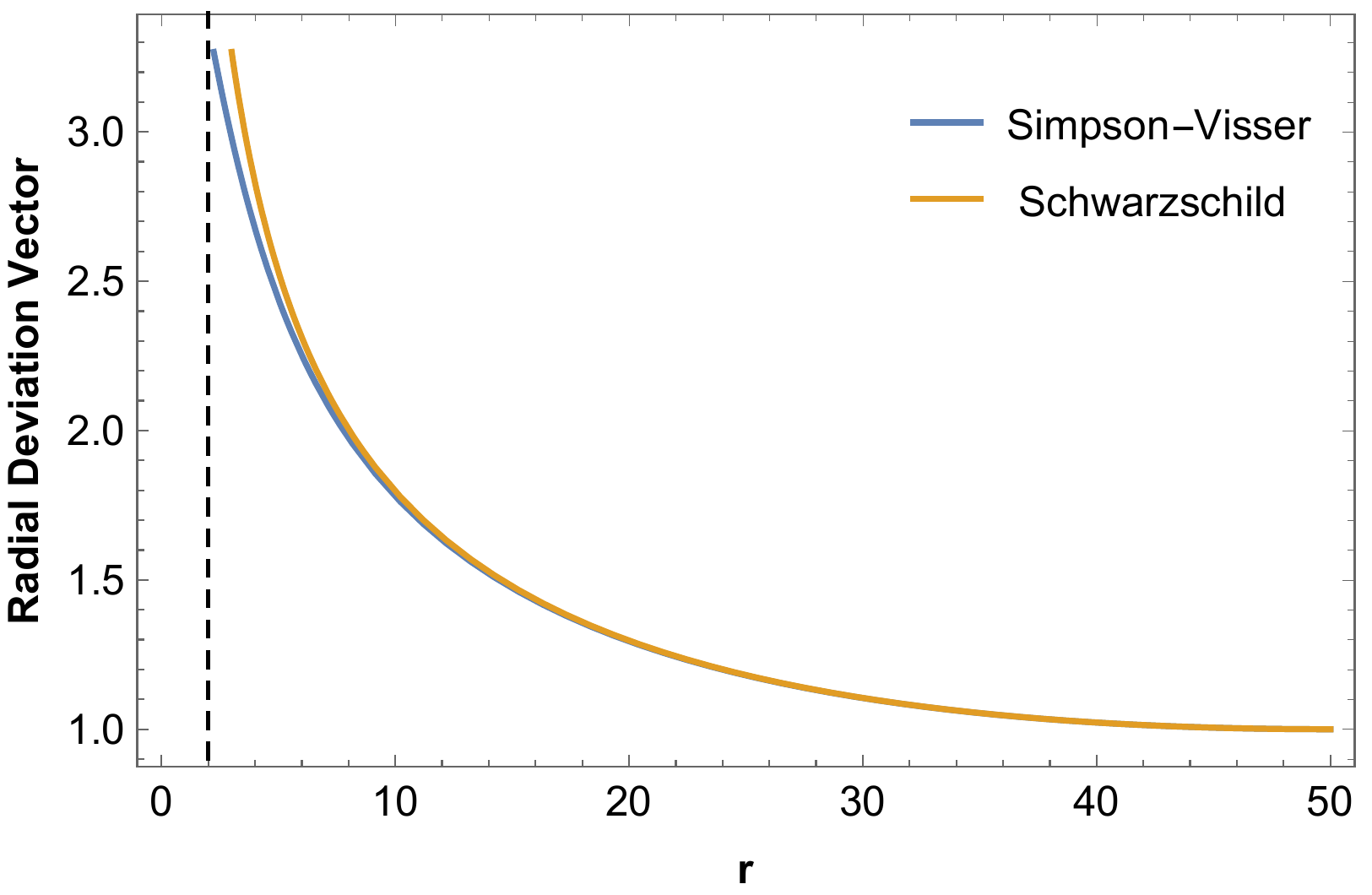}
    
  \end{subfigure}

  \begin{subfigure}{}
    \centering
    \includegraphics[width=\linewidth]{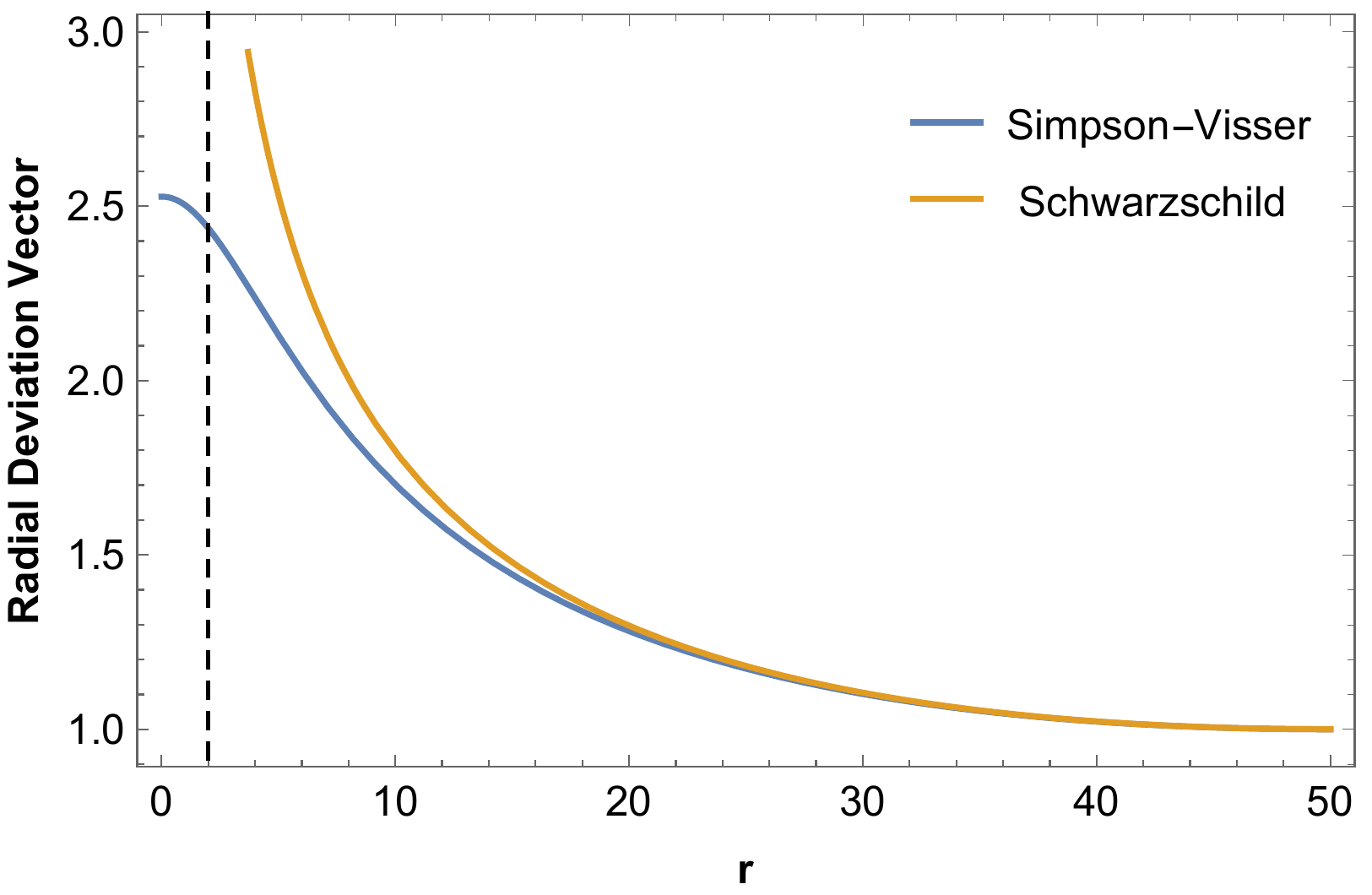}
    
  \end{subfigure}
  \caption{Radial components of geodesic deviation vector as a function of $r$ ($M=1$) using ICI ($b=50M$) for different values of $l$ ($l = 1.8, l=2, l=5$ respectively). Dotted line represents event horizon in Schwarzschild spacetime.}
  \label{fig:6}
  \end{figure}

In Fig.(\ref{fig:6}) we have shown the radial component of the geodesic deviation vector after solving Eq. (\ref{34}) with ICI. We notice that for non zero values for $l$, the radial component of the geodesic deviation vector reaches a finite value as $r \rightarrow 0$, unlike in the Schwarzschild black hole scenario, where it goes till infinity because of the infinite stretching radial tidal force at the singularity. This value keeps on decreasing as the value of $l$ is increased and reaches zero for infinitely large distances. Since the Simpson-Visser spacetime has values of $r$ from $\{-\infty,\infty\}$, the curve shown in the graph is symmetric about the vertical axis.

\begin{figure}[t]
  \centering
  \begin{subfigure}{}
    \centering
    
    \includegraphics[width=\linewidth]{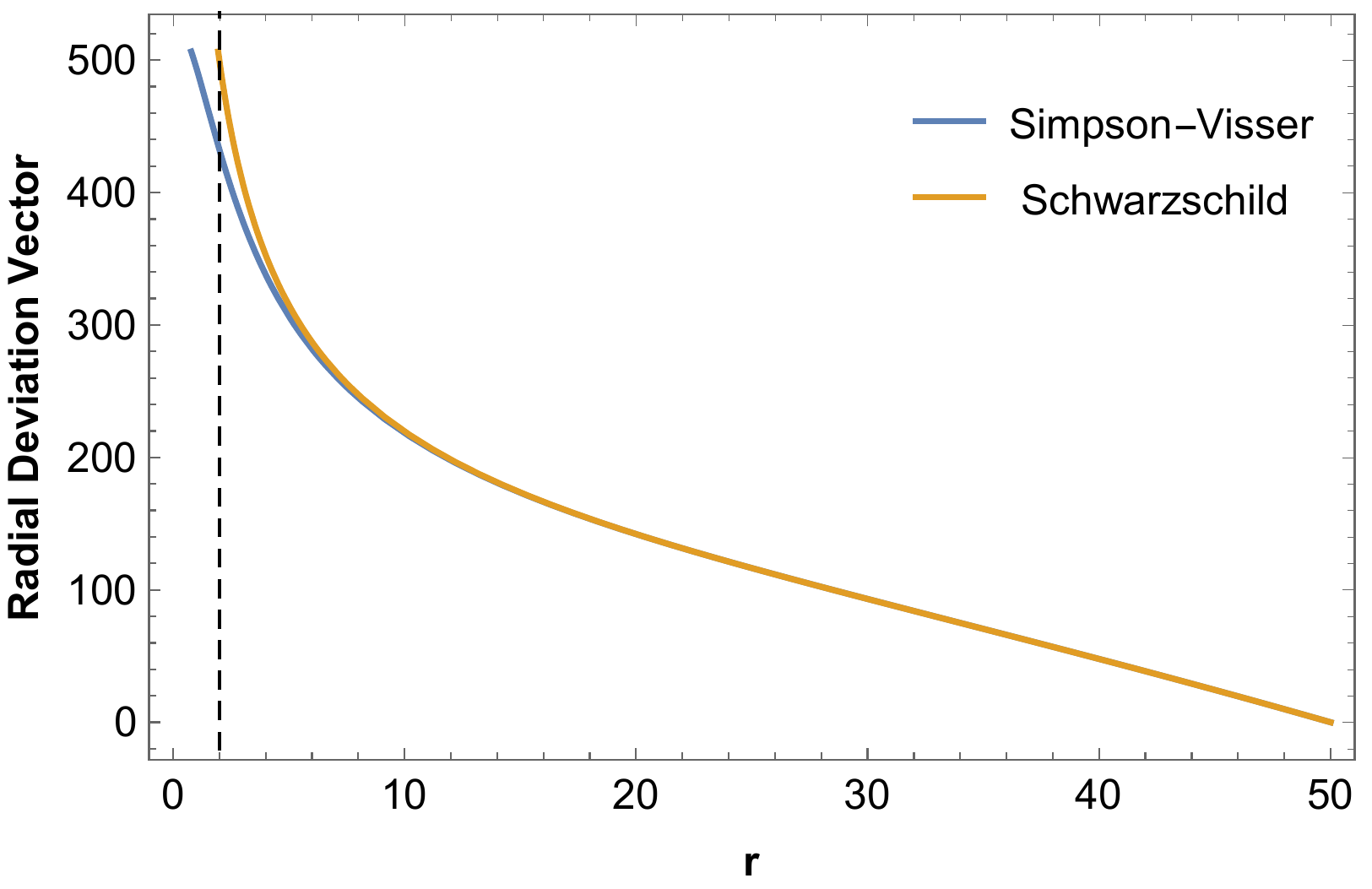}
   
  \end{subfigure}

  \begin{subfigure}{}
    \centering
    \includegraphics[width=\linewidth]{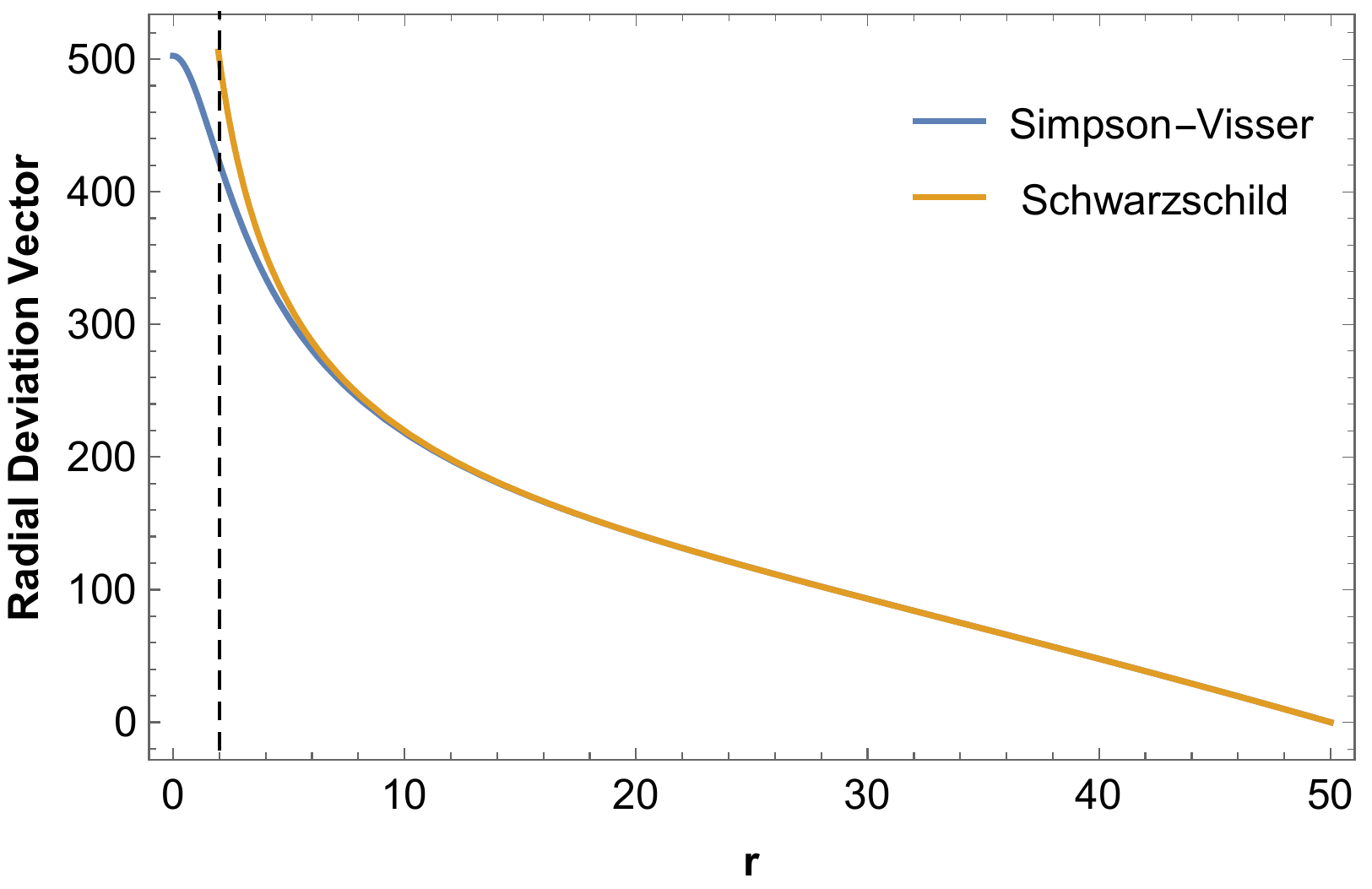}
    
  \end{subfigure}

  \begin{subfigure}{}
    \centering
    
    \includegraphics[width=\linewidth]{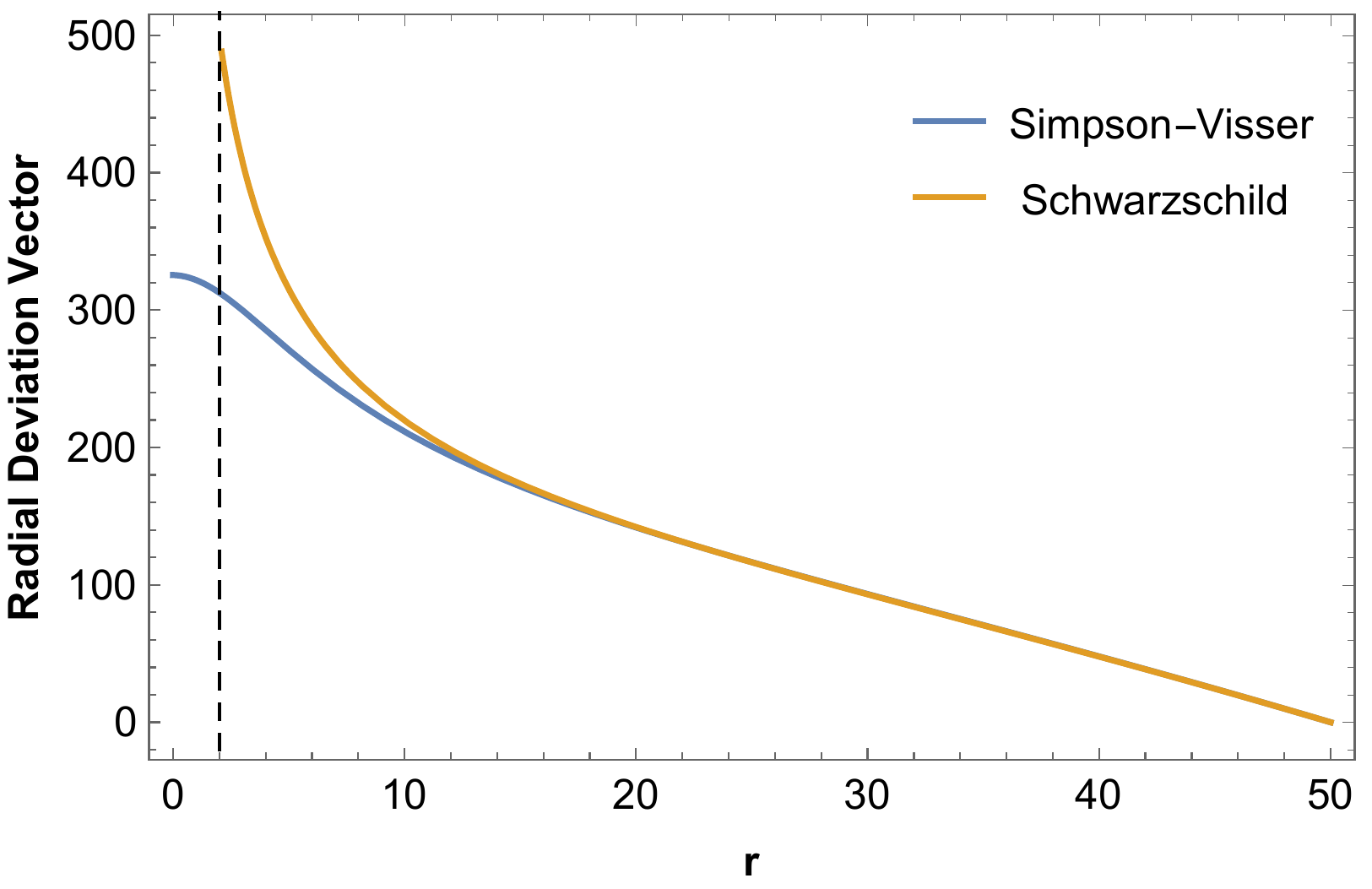}

  \end{subfigure}
  \caption{Radial components of geodesic deviation vector as a function of $r$ ($M=1$) using ICII ($b=50M$) for different values of $l$ ($l = 1.8, l=2, l=5$ respectively). Dotted line represents event horizon in Schwarzschild spacetime.}
  \label{fig:7}
  \end{figure}

By constraining Eq. (\ref{34}) with ICII, we obtain the plot of radial geodesic deviation as shown in Fig.(\ref{fig:7}). It is observed that both the initial conditions show similar trends, but the numerical value of deviation is significantly higher in the case of second initial condition.

\subsection{ANGULAR COMPONENT}\label{V(B)}
In Fig.(\ref{fig:8}) we have shown the angular component of geodesic deviation vector by constraining the Eq. (\ref{35}) with ICI. The results are compared with the Schwarzschild black hole. It is observed that for all values of $l>0$, the value of angular deviation for a radially in-falling particle decreases and reaches a minimum value as $r \rightarrow 0$, and then starts to increase. For very large values of $l$, the angular deviation reaches a constant value. This is in contrast with the Schwarzschild black hole case where the angular deviation reaches $0$ value, as $r \rightarrow 0$. Also, in case of regular black hole geometry ($l<2M$), the point of minima is reached inside the event horizon as shown in fig (\ref{fig:3}). 

For ICII, the angular geodesic deviation profile shows a parabolic trend. It is seen in Fig (\ref{fig:9}) that when a body at rest at some position $r=b$ explodes, the angular deviation starts to increase as the body approaches the center. It reaches a point of maxima and starts to fall rapidly, only to encounter a point of minima which is located outside the horizon for all values of $l<2M$. After crossing this point the angular deviation starts to increase again till $r=0$. The case for Schwarzschild black hole shows a very similar trend except there is no point of minima, and the angular deviation falls to zero after crossing the point of maximum value.

\section{CONCLUSION}\label{VI}
In this paper we investigated the tidal forces in Simpson-Visser spacetime, which comes from the regularization of the Schwarzschild metric. The geometry of this spacetime changes for different values of the regularization parameter $l$, ranging from a regular black hole to a two-way wormhole. We noted that the tidal forces are finite at $r=0$ unlike the Schwarzschild black holes. It is shown that the radial tidal force equation is similar to that of any spherically symmetric and static spacetime, but the angular tidal force equation depends upon the value of the regularization parameter $l$ and energy. We see that the tidal force become zero at a particular value of the radial coordinate which, in the case of radial tidal force solely depends upon the value of $l$. This implies that the radial tidal forces may become compressing instead of stretching and the angular tidal forces may show stretching instead of compressing.

The tidal forces can vanish outside the event horizon for some values of $l$, giving us an opportunity to observe this phenomenon in some cases. We also show that the event horizon for a one-way wormhole ($l=2M$) is formed at $r=0$ and no event horizon is formed in the case of two-way wormhole ($l>2M$). Another important observation made from the radial and angular tidal force curve is that it reaches its peak value later than that in the Schwarzschild black hole case. This shows that the horizon in regular black hole geometries forms after the event horizon in Schwarzschild black hole i.e. $r=2M$. It can also be shown that the angular tidal forces switch their behavior from compression to entirely stretching as the value of $l$ keeps on increasing.

In addition to this, we also examined that the geodesic deviation for a radially falling particle under the influence of tidal forces produced by static regular black hole geometry. Two type of initial conditions are discussed where the first one is when a body is released from rest at a faraway point and the second one is when a body at rest explodes. It is observed that the radial component of geodesic deviation is qualitatively the same for both the initial conditions, but the angular component shows a very peculiar profile, in which it exhibits an oscillatory behavior i.e showing a crest and trough when compared to the Schwarzschild black hole case for different initial conditions, as we showed in section (\ref{V(B)}).
\section{Acknowledgement}
DD would like to acknowledge the support of
the Atlantic Association for Research in the Mathematical Sciences (AARMS) for funding the work.
\clearpage

\clearpage

\begin{figure}[p]
  \centering
  \begin{subfigure}{}
    \centering
    \includegraphics[width=\linewidth]{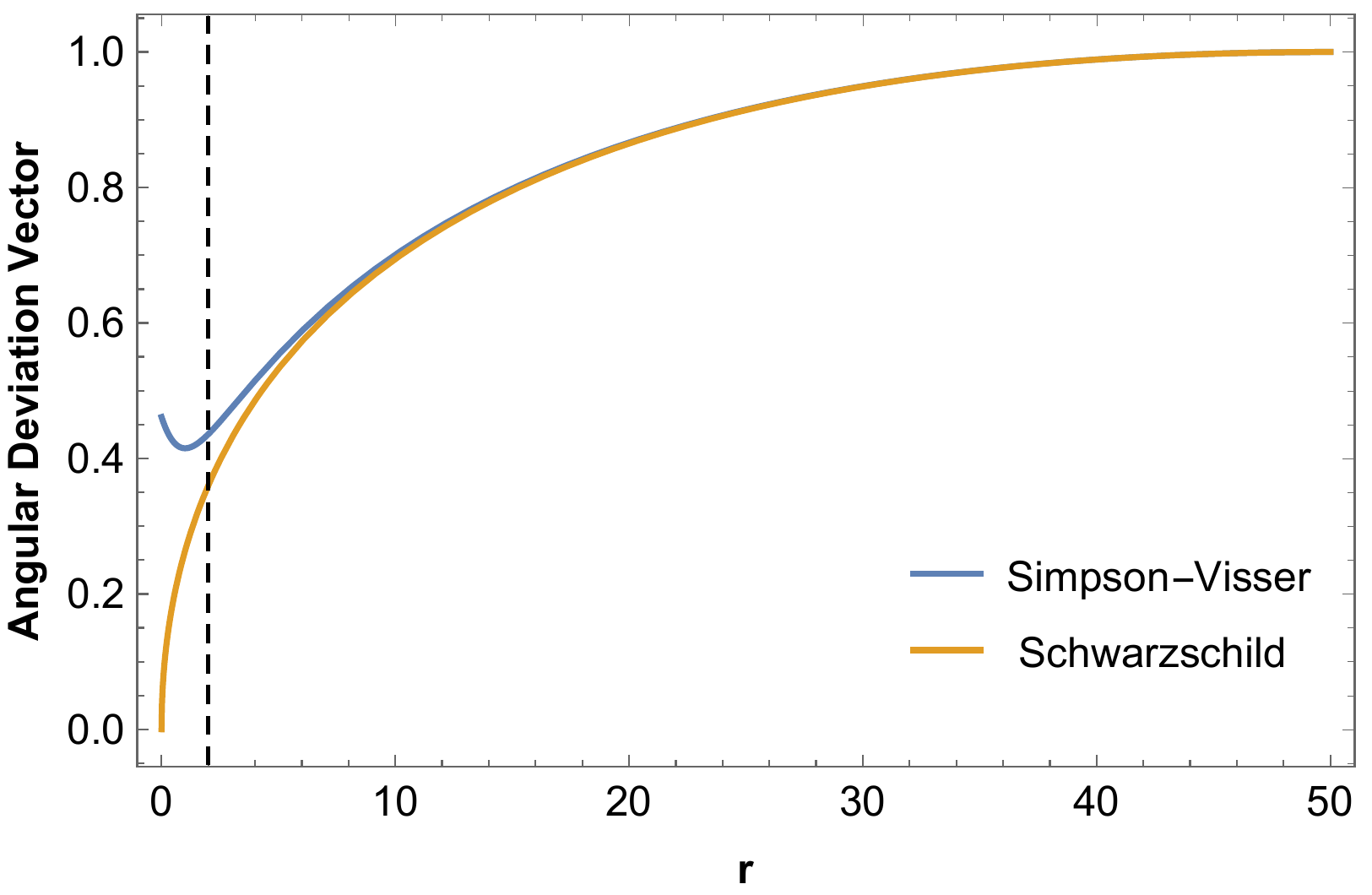}
    
  \end{subfigure}
  \end{figure}

\begin{figure}[p]
  \centering
  \begin{subfigure}{}
    \centering
    \includegraphics[width=\linewidth]{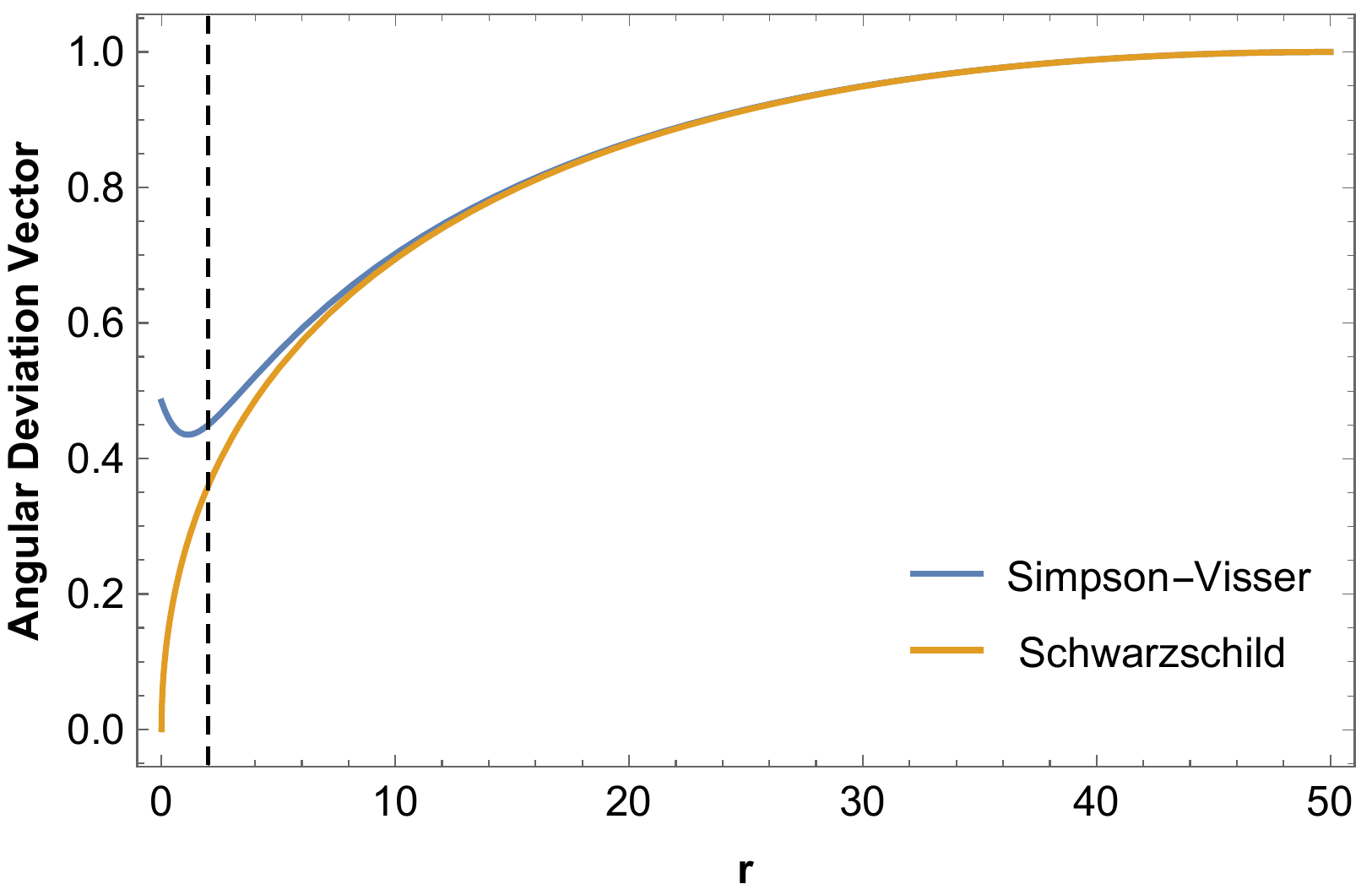}
    
  \end{subfigure}
  \end{figure}

\begin{figure}[p]
  \centering
  \begin{subfigure}{}
    \centering
    \includegraphics[width=\linewidth]{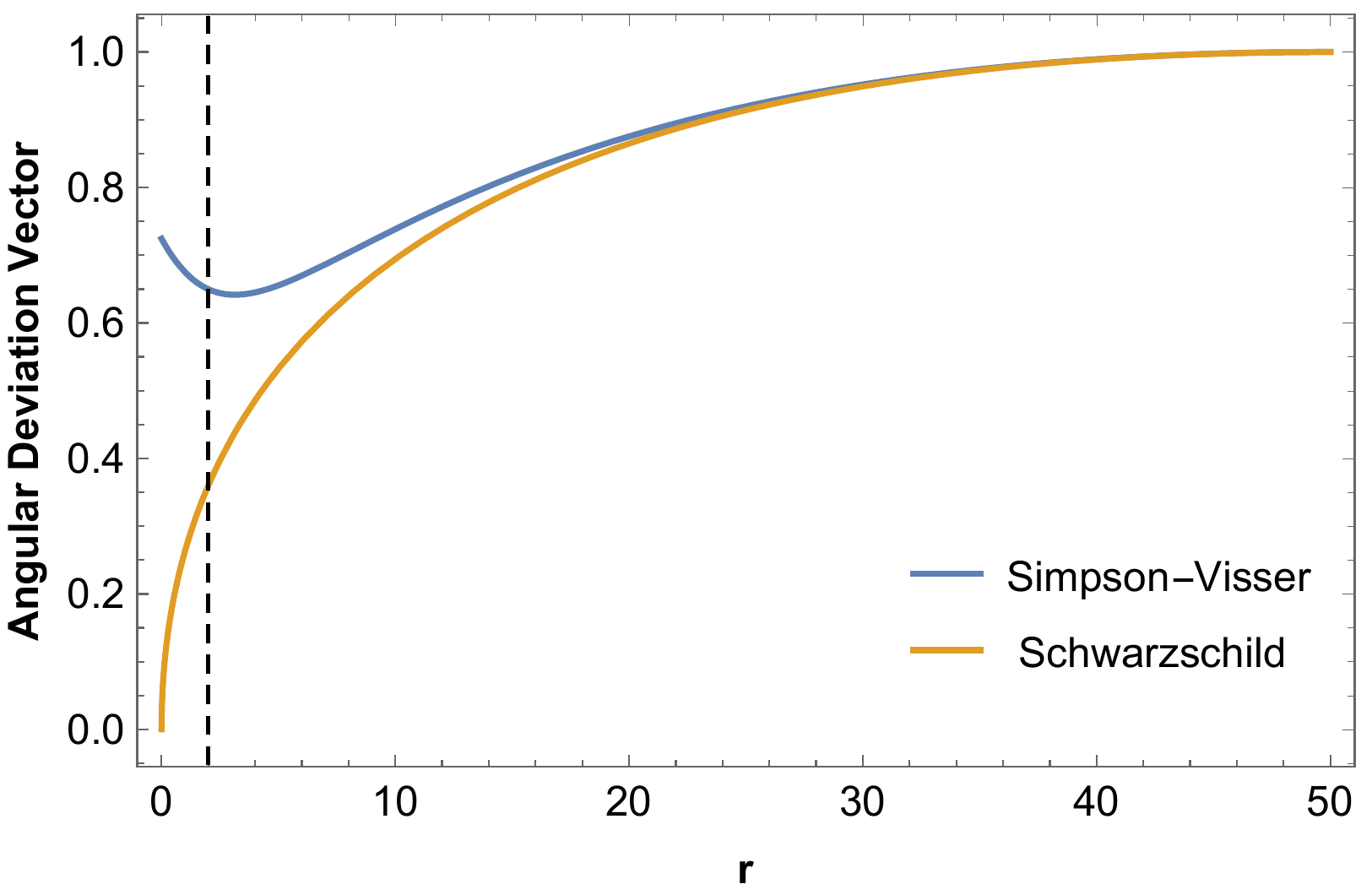}
    \caption{Angular components of geodesic deviation vector as a function of $r$ ($M=1$) using ICI ($b=50M$) for different values of $l$ ($l = 1.8, l=2, l=5$ respectively). Dotted line represents event horizon in Schwarzschild spacetime.}
    \label{fig:8}
  \end{subfigure}
  \end{figure}

\begin{figure}[p]
  \centering
  \begin{subfigure}{}
    \centering
    \includegraphics[width=\linewidth]{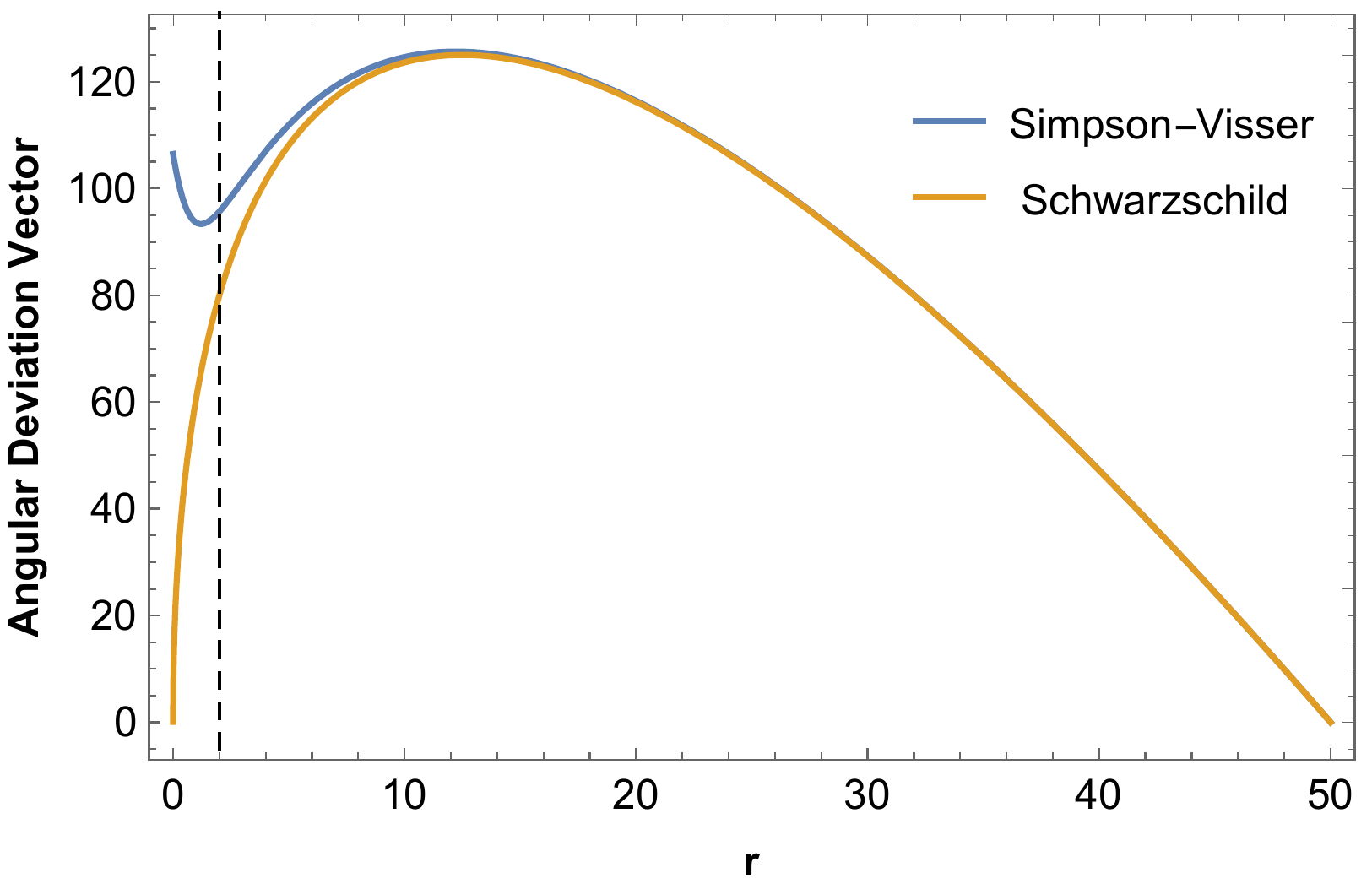}
    
  \end{subfigure}
  \end{figure}

\begin{figure}[p]
  \centering
  \begin{subfigure}{}
    \centering
    \includegraphics[width=\linewidth]{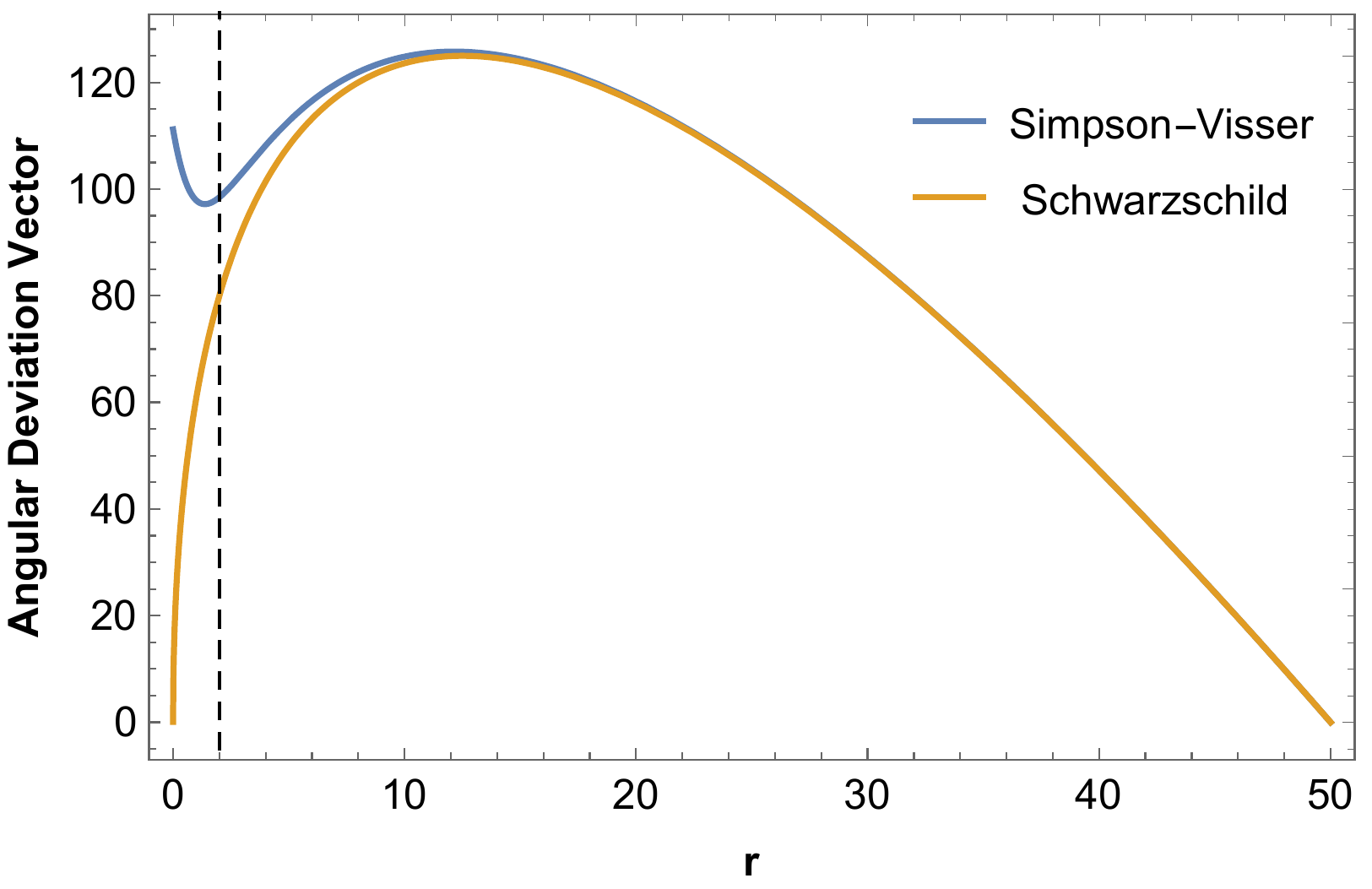}
    
  \end{subfigure}
  \end{figure}

\begin{figure}[p]
  \centering
  \begin{subfigure}{}
    \centering
    \includegraphics[width=\linewidth]{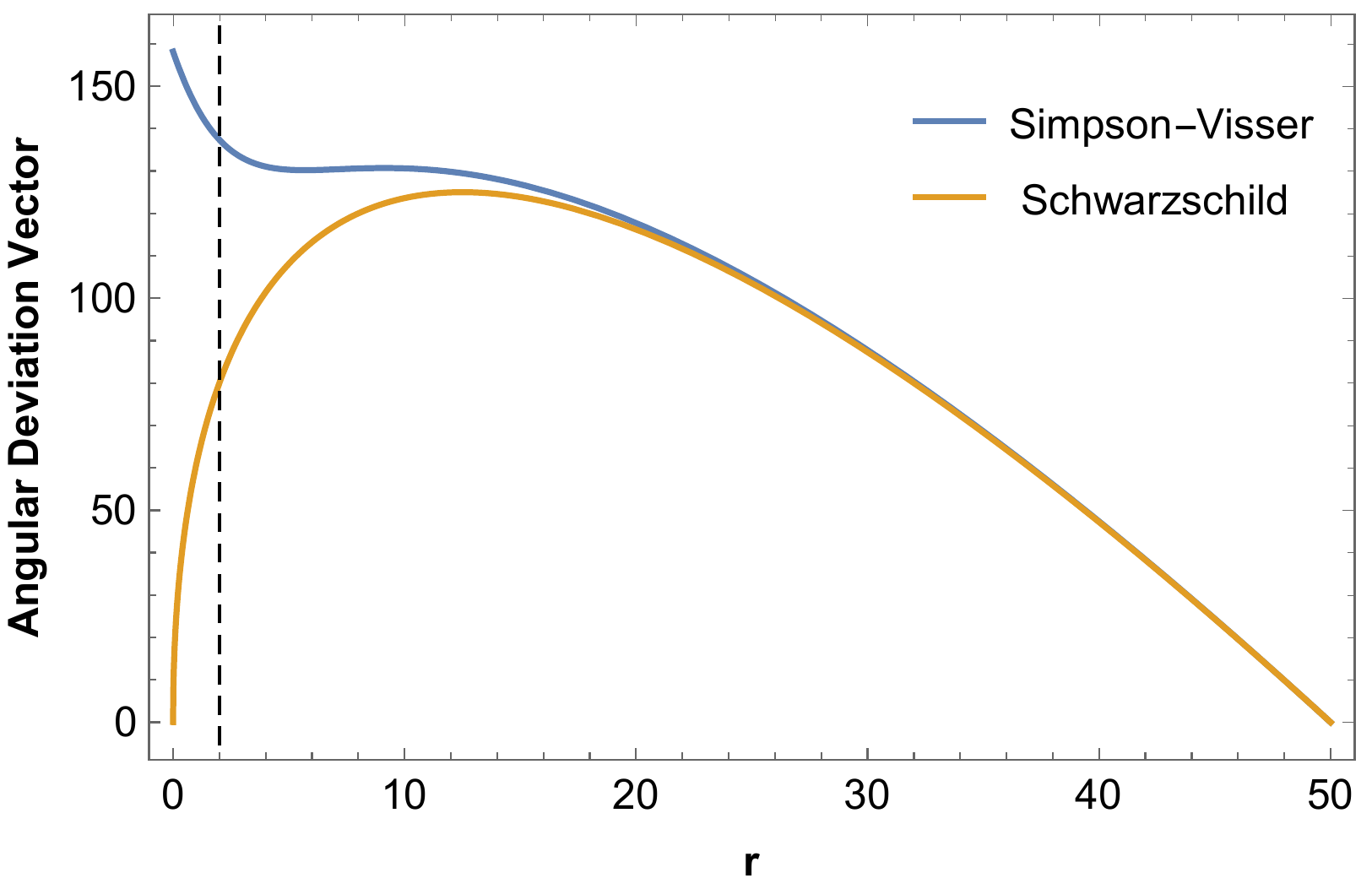}
    \caption{Angular components of geodesic deviation vector as a function of $r$ ($M=1$) using ICII ($b=50M$) for different values of $l$ ($l = 1.8, l=2, l=5$ respectively). Dotted line represents event horizon in Schwarzschild spacetime.}
    \label{fig:9}
  \end{subfigure}
  \end{figure}

\end{document}